\documentclass[aps,prl,twocolumn, showpacs]{revtex4}
\usepackage{graphicx}
\usepackage{amsmath}
\usepackage{amssymb}
\usepackage{bm}
\usepackage{dcolumn}
\usepackage{color}
\usepackage{hyperref}

\begin{document}

\title{Quantum Metrological Limits via a Variational Approach}

\author{B. M. Escher}\email{bmescher@if.ufrj.br} \affiliation{Instituto de F\'{\i}sica, Universidade Federal do Rio de Janeiro, 21.941-972, Rio de Janeiro (RJ) Brazil}
\author{L. Davidovich} \affiliation{Instituto de F\'{\i}sica, Universidade Federal do Rio de Janeiro, 21.941-972, Rio de Janeiro (RJ) Brazil}
\author{N. Zagury} \affiliation{Instituto de F\'{\i}sica, Universidade Federal do Rio de Janeiro, 21.941-972, Rio de Janeiro (RJ) Brazil}
\author{R. L. de Matos Filho} \affiliation{Instituto de F\'{\i}sica, Universidade Federal do Rio de Janeiro, 21.941-972, Rio de Janeiro (RJ) Brazil}

\date{\today}

\begin{abstract}
The minimum achievable statistical uncertainty in the estimation of physical parameters is determined by the quantum Fisher information. Its computation for noisy systems is still a challenging problem. Using a variational approach, we present an equation for obtaining the quantum Fisher information, which has an explicit dependence on the mathematical description of the noise. This method is applied to obtain a useful analytical bound to the quantum precision in the estimation of phase-shifts under phase diffusion, which shows that the estimation uncertainty cannot be smaller than a noise-dependent constant.
\end{abstract}


\maketitle

\textit{Introduction}.---Quantum metrology~\cite{Helstrombook76,Holbook82,Braunsprl94,Giosci04} deals with the estimation of parameters taking into account the constraints imposed by quantum  laws. The estimation is based on measurements made on probe systems undergoing a parameter-dependent process. For a given measurement scheme, the uncertainty  in the estimation of a parameter is limited by the Cram\'er-Rao bound, which is proportional to the inverse of the square root of the so-called Fisher information (FI)~\cite{fisher22,Cramerbook46,Raobook73}. The maximization of  FI over all measurement strategies allowed by quantum mechanics leads to a non-trivial quantity: the quantum Fisher information (QFI). The determination of this quantity is central to quantum metrology. It allows, for instance, the establishment of ideal benchmarks for the statistical uncertainty  in the estimation of parameters, which can be used by experimentalists to evaluate the performance of a real experiment. A systematic approach to calculate the QFI, using the symmetric logarithmic derivative (SLD) operator~\cite{Helstrombook76,Holbook82},  was developed in Ref.~\cite{Braunsprl94}. This  approach has allowed large advances on quantum metrology~\cite{Gionpho11,Bananpho09}. For unitary processes, it leads to simple analytical expressions. This is not the case, however, for noisy processes, which often require numerical calculations. 

When the unknown parameter is associated with a physical process, the ultimate limit for the precision in its estimation is given by a further maximization of the QFI over all initial probe  states (given some constraint, e. g. a fixed average energy). These two maximizations make the determination of that ultimate limit a laborious numerical task. Recently, an alternative to solve this problem was presented in Ref.~\cite{Eschernp11}: given a mathematical description of the quantum parameter-dependent process by a set of  Kraus operators~\cite{Krausbook83}, an upper bound to the QFI can be calculated; the true value of the QFI is obtained by minimizing this upper bound over all equivalent Kraus representations of the process.
  
In this Letter, we present a variational approach, based on purification techniques~\cite{Nielsenbook01}, to calculate the QFI through the minimization of upper bounds. These upper bounds correspond  to the QFI associated with pure states in the enlarged Hilbert space of all purifications of the original probe state. An important advantage of our approach is that it results in a general prescription for performing that minimization: beginning with an arbitrary purification of the probe state, the optimum purification that minimizes these upper bounds and yields the QFI can be found through the solution of a Sylvester equation \cite{Bartels72}. Such a prescription, which leads to an alternative way of expressing the quantum Fisher information, actually solves the minimization problem posed in  Ref.~\cite{Eschernp11}. As a concrete example, we use this approach to delimit the ultimate precision bounds on the estimation of phase-shifts in the presence of phase-diffusion. This problem was addressed recently in~\cite{Parisprl11}, where these bounds were found numerically for initial Gaussian probe states, and posteriorly confirmed experimentally in the special case of initial coherent states~\cite{Parisexp}. Here, we show an analytical and nontrivial lower bound for this quantum limit, which is valid for any probe state. This bound reveals a drastic effect of phase-diffusion on phase estimation: the accuracy, even though dependent on the energy of the probe state,  cannot be better than a noise-dependent constant.

\textit{QFI by a variational approach}.---The estimation of a parameter is based on experimental results of measurements on a quantum probe state, after it has been submitted to a physical operation that depends on the value of the parameter. Usually, it is previously assumed that the possible values of the parameter to be estimated, denoted here by $x$, lie within a certain continuous interval. It is also assumed that one knows the precise dependence of the physical operation on the value of the parameter $x$, so that, for a given initial probe state, one knows the transformed state $\hat{\rho}(x)$. Finally, one assumes  a specific measurement device, mathematically represented by positive operator-valued measures $\{\hat{E}_{k}\}$~\cite{Nielsenbook01}. The goal, therefore, is to obtain an accurate estimation, $x_{\textrm{est}}(k)$, of the true value $x_{\textrm{true}}$ of the unknown parameter, from a set of experimental results $k$, using a given rule. The variance between an estimate and any possible value of the parameter is $\delta x \equiv \sqrt{\langle(x_{\textrm{est}}- x)^2\rangle}$, where $\langle\bullet\rangle\equiv\sum_{k}p_{k}(x)\bullet$ and $p_{k}(x)=\textrm{Tr}[\hat{\rho}(x)\hat{E}_{k}]$. Here, $p_{k}(x)$ is the probability of obtaining the set of experimental results $k$ given that the parameter value is $x$. This variance may be considered a merit quantifier for the estimation as a function of $x$. For any unbiased estimation (i.e., $\langle x_{\textrm{est}}\rangle=x$), the statistical uncertainty is limited by the Cram\'{e}r-Rao bound~\cite{Cramerbook46,Raobook73}. For an experiment with $\nu$ repetitions, this bound is given by $\delta x \ge 1/\sqrt{\nu F(x)}$, where $F(x)=\sum_{k}p_{k}(x)[d\ \textrm{ln}[p_{k}(x)]/dx]^2$ is the FI. Under very general assumptions~\cite{fisher22,Cramerbook46,Raobook73}, it can be shown that it is possible to saturate the Cram\'{e}r-Rao bound, at least in the asymptotic regime ($\nu\rightarrow\infty$).

The QFI is defined by the maximum of the FI over all possible measurement strategies allowed by quantum physics:
\begin{equation}
{\cal F}_{Q}[\hat{\rho}(x)]=\max_{\{\hat{E}_{k}\}} F[\hat{\rho}(x);\{\hat{E}_{k}\}] \ .
\end{equation}
The respective quantum version of the Cram\'er-Rao inequality~\cite{Milburn96}, $\delta x\ge 1/\sqrt{\nu {\cal F}_{Q}(x)}$, settles, therefore, a limit to the statistical uncertainty that cannot be overcome by any strategy of estimation, for a given physical process and quantum probe.

If the transformed probe state is pure, $\hat{\rho}(x)=\vert\psi(x)\rangle\langle\psi(x)\vert$, the correspondent expression of the QFI is \cite{Helstrombook76,Holbook82,Braunsprl94,Escherbjp11}:
\begin{equation}\label{dfq}
{\cal F}_{Q}[\hat\rho(x)]\!=\!4\left[\!\dfrac{d\langle\psi(x)\vert}{dx}\!\dfrac{d\vert\psi(x)\rangle}{dx}\!-\!\left\vert\dfrac{d\langle\psi(x)\vert}{dx}\!\vert\psi(x)\rangle\!\right\vert^{2}\right].
\end{equation}

If the state $\hat{\rho}(x)$ is not pure, the SLD approach does not lead in general to such a simple analytical expression. On the other hand, it is always possible to enlarge the size of the original Hilbert space $S$ and build a pure state $\vert\Phi_{S,E}(x)\rangle$ in the enlarged space $S+E$ that fulfills the  condition ${\rm Tr}_E\hat{\rho}_{S,E}(x)=\hat{\rho}_{S}(x)$, where $\hat\rho_{S,E}(x)=\vert\Phi_{S,E}(x)\rangle\langle\Phi_{S,E}(x)\vert$, and the trace is taken only on the $E$-Hilbert space~\cite{Nielsenbook01}. We have added the label $S$ to the state $\hat\rho(x)$ of the system, in order to distinguish it from states in space $S+E$. The state $\vert\Phi_{S,E}(x)\rangle$ is called a purification of $\hat{\rho}_{S}(x)$ and the space $E$ may be interpreted as the Hilbert space corresponding to an environment for system $S$.

Because taking the trace over $E$ may be viewed as discarding information on part of the total space $S+E$, a physically motivated upper bound $C_{Q}$ of ${\cal F}_{Q}[\hat{\rho}_S(x)]$ can be obtained:
\begin{equation}
C_{Q}\left[\hat\rho_{S,E}(x)\right]\equiv{\cal F}_{Q}[\hat\rho_{S,E}(x)] \ge{\cal F}_{Q}[\hat{\rho}_S(x)]\, ,
\end{equation}
this inequality being valid for any purification of $\hat{\rho}_S(x)$. Physically, this is due to the fact that when a system plus an environment are monitored together, the information obtained about an unknown parameter cannot be smaller than the information acquired when only the system is measured. Since $C_{Q}$ depends on the purification chosen, the best upper bound that can be obtained with this strategy is given by the minimum of $C_{Q}$ over all possible purifications of $\hat\rho_{S}(x)$. In the supplementary material of Ref.~\cite{Eschernp11}, it is shown that this minimization can be performed on the restricted set of all purifications of $\hat\rho_S(x)$ belonging to a given space $S+E$, as long as the dimension of $E$ is at least equal to the dimension of $S$, with this minimum being equal to the QFI: 
\begin{equation}\label{cqmbr}
{\cal F}_{Q}[\hat{\rho}_{S}(x)]\equiv\min_{\{\vert\Phi_{S,E}(x)\rangle\}}C_{Q}\left[\hat\rho_{S,E}(x)\right] \ .
\end{equation} 

\textit{The minimization of $C_{Q}$}.---It is possible to determine the value of QFI by minimizing the upper bound $C_{Q}$ over all purifications of $\hat{\rho}_{S}(x)$ in a given enlarged state space $S+E$ \cite{Eschernp11}. This is, in general, a challenging task and a concrete prescription to do it would be welcome~\cite{Gionp11}. In the following we present such a prescription. Our procedure starts by establishing a relation between all purifications in a given space $S+E$. As shown in Ref.~\cite{ Nielsenbook01}, there is always a unitary operator $\hat{u}_{E}(x)$, acting effectively only on the $E$ space, that connects two purifications $\vert\Psi_{S,E}(x)\rangle$ and $\vert\Phi_{S,E}(x)\rangle$ of the same state $\hat{\rho}_{S}(x)$:
\begin{equation}\label{pubr}
\vert\Psi_{S,E}(x)\rangle=\hat{u}_{E}(x)\vert\Phi_{S,E}(x)\rangle \ ,
\end{equation}
where $\hat u_{E}(x)$ is a shorthand for the operator $\hat{u}_{E}(x)\otimes\openone_{S}$, and $\openone_{S}$ is the identity operator on space $S$. Therefore, given a purification $\vert\Phi_{S,E}(x)\rangle$, the QFI may be found by minimizing $C_{Q}[\hat{u}_{E}(x)\hat\rho_{S,E}(x)\hat{u}_{E}^\dagger(x)]$ over all unitary operators on $E$ space. The physical role of $\hat{u}_{E}(x)$ is to erase all nonredundant information about the parameter $x$ that has been leaked from space $S$ into the larger space $S+E$. 

It will be useful to define a Hermitian operator $\hat{h}_{E}(x)$, which acts effectively only in the $E$ space, by
\begin{equation}
\hat{h}_{E}(x)=i\dfrac{d\hat{u}^{\dagger}_{E}(x)}{dx}\hat{u}_{E}(x) \ ,
\end{equation}   
and another Hermitian operator $\hat{H}_{S,E}(x)$, which acts in the whole $S+E$ space, by
\begin{equation}
i\dfrac{d\vert\Phi_{S,E}(x)\rangle}{dx}=\hat{H}_{S,E}(x)\vert\Phi_{S,E}(x)\rangle \ .
\end{equation} 
Using the definitions above, we may write $C_{Q}$ as
\begin{equation}\label{cqh}
{C_Q}=4\langle[\hat{{\cal H}}(x)-\langle\hat{{\cal H}}(x)\rangle_\Phi] ^2\rangle_\Phi  \ ,
\end{equation}
where $\hat{\cal H}(x)=\hat{H}_{S,E}(x)-\hat{h}_{E}(x)$, and the averages are taken over $\vert\Phi_{S,E}(x) \rangle$. From Eq.~(\ref{cqh}), we conclude that the minimization of $C_Q$ over all unitary operators $\hat{u}_{E}(x)$ is equivalent to the minimization of ${C_Q}$ over all Hermitian operators $\hat{h}_{E}(x)$ that act on $E$ space.
This minimization is a mathematical optimization problem in positive semidefinite quadratic programing, which can be efficiently solved numerically~\cite{Vanbook09}, since the operator $\hat{h}_{E}(x)$ appears as a quadratic function in $C_Q$. Thereupon, it is possible to find an equation for the optimum Hermitian operator $\hat{h}_{E}^{\rm{(opt)}}(x)$ that minimizes ${C_{Q}}$. Taking, without loss of generality, $\langle\hat{h}_{E}^{\rm{(opt)}}(x)\rangle_{\Phi}=\langle\hat{H}_{S,E}(x)\rangle_{\Phi}$, one finds that:
\begin{equation}\label{hopt}
\dfrac{\hat{h}_{E}^{\rm{(opt)}}\!(x)\hat{\rho}_{E}(x)+\hat{\rho}_{E}(x)\hat{h}_{E}^{\rm{(opt)}}\!(x)}{2}\!=\!{\rm Tr}_S\!\!\left\{{\cal D}\!\left[\hat{\rho}_{S,E}(x)\right]\right\}\!,
\end{equation}
where $\hat{\rho}_{E}(x)=\rm{Tr}_{S}[\vert\Phi_{S,E}(x)\rangle\langle\Phi_{S,E}(x)\vert]$ is the reduced density matrix in the $E$-space, and ${\cal D}\left[\hat\rho_{S,E}(x)\right]$ is defined as:
\begin{equation}
{\cal D}\!\left[\hat\rho_{S,E}(x)\right]\!\equiv\!\dfrac{i}{2}\left[\dfrac{d\vert\Phi_{S,E}\rangle}{dx}\langle\Phi_{S,E}\vert-\vert\Phi_{S,E}\rangle\dfrac{d\langle\Phi_{S,E}\vert}{dx}\right]\!.
\end{equation}
Equation (\ref{hopt}), when expressed in terms of the matrices associated to the corresponding operators, is a Sylvester equation~\cite{Bartels72}. It depends only on the degrees of freedom of $E$. After determining $\hat{h}_{E}^{\rm{(opt)}}(x)$, the QFI may be finally expressed as
\begin{equation}
{\cal F}_{Q}[\hat{\rho}_S(x)]=C_{Q}[\hat\rho_{S,E}(x)] - 4\langle[\Delta\hat{h}_{E}^{\rm (opt)}(x)]^2\rangle_\Phi \ .
\end{equation}
This is a novel expression for the quantum Fisher information, which relates it directly to the QFI corresponding to a unitary evolution of the enlarged system, and shows that the nonredundant information in $\vert\Phi_{S,E}(x)\rangle$ about the parameter $x$ is given by four times the variance of $\hat{h}_{E}^{\rm (opt)}(x).$ As compared to the expression for the QFI of the system alone, given by the SLD approach, it displays explicitly the mathematical description of the noise process, through the purification $\vert\Phi_{S,E}(x)\rangle.$

The usefulness of this method is not restricted to obtaining an equation for the exact evaluation of QFI. Since our approach is variational, whenever it is too hard to find a solution of Eq.~(\ref{hopt}), we may still  obtain satisfactory and nontrivial analytical upper bounds to QFI. Indeed, based on (\ref{hopt}), one may guess approximations for $\hat h_{E}^{\rm (opt)}(x)$ that depend on variational parameters, so that the minimization is made on subclasses of operators $\hat h_{E}(x)$. This procedure leads to bounds for QFI, given by the minima of (\ref{cqh}) over these subclasses. It also allows an iteration procedure for getting progressively better approximations to the QFI of the system.

\textit{Phase estimation under phase-diffusion}.---The estimation of phase shifts is a central problem in quantum optics, metrology and quantum communication~\cite{Parisprl11,Parisexp,geral}. It is important, for example, in the use of light interferometers as part of detectors of gravitational waves~\cite{ligo11}. When planning such experiments, it is essential to take into account the unavoidable influence of noise on the ultimate precision limits. Besides photon losses,  phase diffusion is another relevant source of noise in optical phase measurements and must be taken into account. For incoming Gaussian states, a numerical study of the effect of phase diffusion on the ultimate limit of precision for phase estimation was presented in~\cite{Parisprl11}. 

In the following, we apply our approach to this problem in order to derive nontrivial analytical lower bounds to this limit, which are valid for any probe state. For concreteness, consider an initially pure probe state $\hat{\rho}_{S}=\vert\psi_{S}\rangle\langle\psi_{S}\vert,$  corresponding to a generic harmonic oscillator, which undergoes a phase shift $\phi$ due to some physical process. The resulting state of the probe, in the presence of phase diffusion noise, may be represented, in the Markov limit, by
\begin{equation}
\hat{\rho}_{S}(\phi)=\sum_{m,n=0}^\infty\rho_{m,n}e^{-i\phi(m-n)-\beta^{2}(m-n)^2}\vert m_S\rangle\langle n_S\vert \ ,
\end{equation}
where $\rho_{m,n}$ is the matrix element of the initial probe state in the Fock basis of the system and $\beta$ quantifies the degree of diffusion present in the process (from $\beta=0,$ corresponding to no diffusion, to $\beta=\infty$, corresponding to maximum diffusion). In order to obtain a possible purification of $\hat\rho_{S}(\phi)$ based on physical insight, we consider an optical interferometer with a dispersive plate producing a difference of phase $\phi$ between its two arms, and model the phase diffusion of the initial probe state  through the effect of the radiation pressure on one of the interferometer mirrors. The interaction between the light field and the mirror is taken as proportional to $\hat{n}_{S}\hat{x}_{E}$, where $\hat n_S$ is the photon number operator and $\hat{x}_{E}$ is the dimensionless position operator of the mirror. In this model, the final state of the combined system of the probe and the mirror is given by 
\begin{equation}
\vert\Phi_{S,E}(\phi)\rangle=e^{-i\phi\hat{n}_{S}}e^{i(2\beta)\hat{n}_{S}\hat{x}_{\rm{E}}}\vert\psi_{S}\rangle\vert 0_{E}\rangle \ ,
\end{equation}
where $\vert 0_{E}\rangle$ is the initial state of the mirror, which we assume to be the ground state of a quantum oscillator. It is straightforward to see that this state is, indeed, a purification of $\hat\rho_S(\phi)$. The value of $C_{Q}[\hat\rho_{S,E}(\phi)]$ may be now calculated directly through Eq.~(\ref{dfq}): $C_{Q}=4\Delta n^{2}$, where $\Delta  n^{2}$ is the variance of the photon number operator in the initial probe state. Notice that this purification leads to a trivial upper bound to the QFI of $\hat\rho_{S}(\phi)$ as it is equal to the QFI in the absence of phase diffusion ($\beta=0$). 

In order to obtain a tighter upper bound to the QFI of $\hat\rho_{S}(\phi)$, we consider possible approximations to $\hat h_{E}^{\rm (opt)}(\phi)$ by analyzing more closely Eq.~(\ref{hopt}). The reduced density matrix of the mirror associated with the purification $\vert\Phi_{S,E}(\phi)\rangle$ is 
\begin{equation}
\hat\rho_{E}=\sum_{n=0}^\infty\vert \rho_{n,n}\vert^2\vert i\sqrt{2}\beta{n}\rangle_{E}\langle i\sqrt{2}\beta{n}\vert \, ,
\end {equation} 
where $\vert i \sqrt{2}\beta{n}\rangle_{E}$ is a coherent state with amplitude $\sqrt{2}\beta{n}$. The right-hand side of Eq.~(\ref{hopt}) is ${\rm Tr}_{S}{\cal D}[\hat\rho_{S,E}]=[-i\hat{b}_{E}/(2\sqrt{2}\beta)]\hat\rho_{E}+i\hat\rho_{E}[\hat{b}^{\dagger}_{E}/(2\sqrt{2}\beta)]$, where $\hat{b}_{E}=(\hat{x}_{E}+i\hat{p}_{E})/\sqrt{2}$, with $\hat{p}_{\rm{E}}$ being the dimensionless momentum operator of the mirror. Notice that the solution of Eq.~(\ref{hopt}) for $\hat h_{E}^{\rm (opt)}(\phi)$ would be trivial if $i\hat b_E$ were an hermitian operator. However, in the asymptotic regime, $\sqrt{2}\beta{n}\gg{1},$ the operator $\hat{p}_{E}/(2\beta)$ when applied to $\hat\rho_E$ produces a result  quite similar to that of $-i\hat b/(\beta\sqrt{2}).$ So, we may guess that $\hat{u}_{E}(\phi;\lambda)=e^{i\phi\lambda\hat{p}_{\rm{E}}/(2\beta)}$ (with $\lambda$ being a variational parameter) would be a reasonable candidate to erase part of nonredundant information in $\vert\Phi_{S,E}(\phi)\rangle.$ In this case the upper bound of QFI is:
\begin{equation}
C_{Q}=(1-\lambda)^{2}4\Delta{n}^2 + \lambda^{2}/(2\beta^2) \ .
\end{equation}
The optimal value of $\lambda$ that minimizes $C_{Q}$ is $\lambda_{\rm{opt}}=8\Delta{n}^2\beta^2/(1+8\Delta{n}^2\beta^2)$. Then, taking the inverse of the square root of $C^{\rm opt}_{Q}$, one gets a nontrivial bound for the precision of phase-shift estimation in the presence of phase diffusion, valid for any input state:
\begin{equation}\label{lb}
\delta\phi\ge\sqrt{\dfrac{1}{\nu C_{Q}^{\rm opt}}}=\sqrt{\dfrac{1}{\nu}\left(\dfrac{1}{4\Delta{n}^{2}}+ 2\beta^2\right)} \ .
\end{equation}
This inequality shows that the uncertainty in this estimation is limited by a well-known formula for independent noise sources, displaying clearly the effects of the intrinsic probabilistic feature of quantum mechanics, $1/(4\Delta n^2),$ and of the genuine phase-diffusion noise, $2\beta^2$. 

\begin{figure}[t]
\centering
\includegraphics[width=0.45\textwidth]{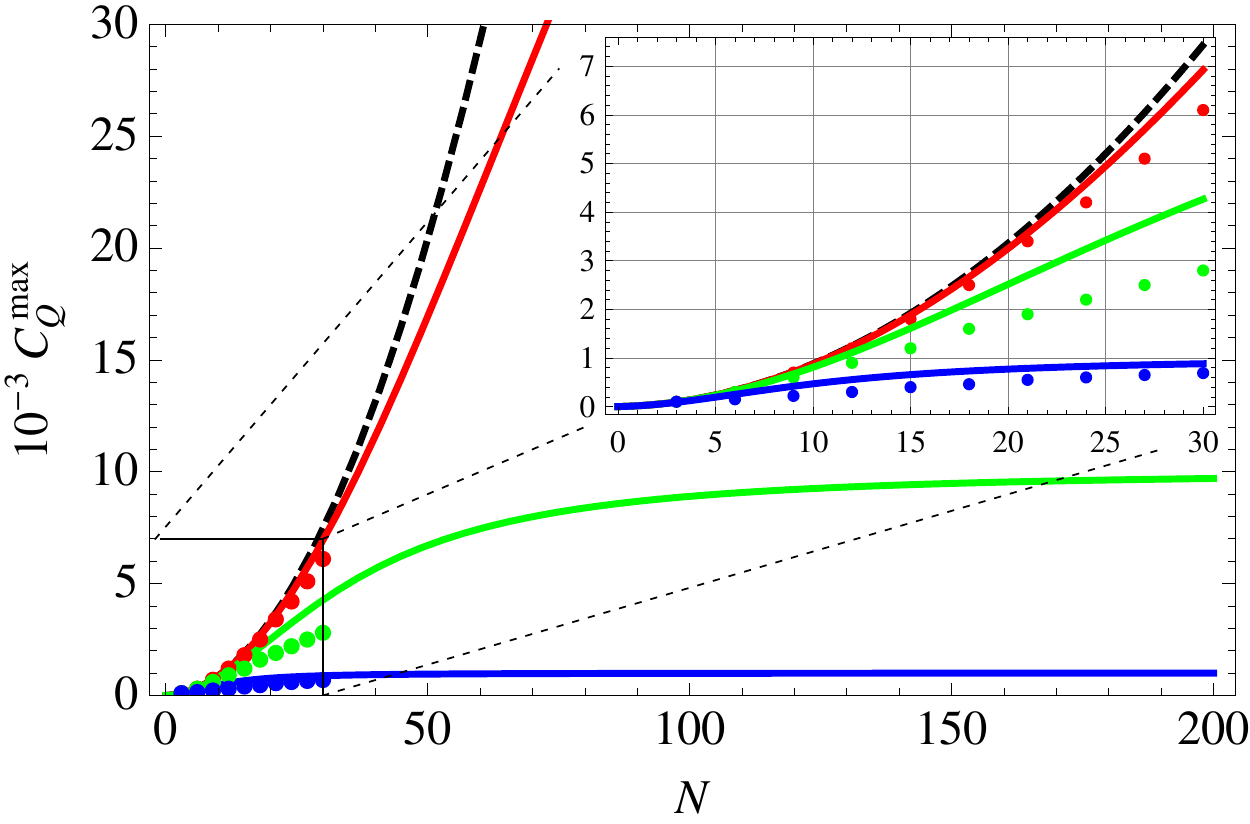}
\caption{Comparison between upper bound $C_Q^{\rm max}$ and the maximum quantum Fisher information ${\cal F}_Q^{\rm max}$  in~\cite{Parisprl11} as a function of the average number of photons $N$. The dots stand for the values obtained in~\cite{Parisprl11}, the dashed line corresponds to the noiseless case ($\beta^2=0$), and the full lines correspond to $C_Q^{\rm max}$. The inset displays the two quantities up to $N=30$, which was the range considered in~\cite{Parisprl11}. From bottom to top, $\beta^2=5\times10^{-4}; 5\times 10^{-5};5\times10^{-6}.$}
\label{fig1}
\end{figure}

An important property of the bound shown above is the presence of a constant term. This means that the presence of phase diffusion is, in general, more detrimental to phase-shift estimation than the presence of photon losses, when the uncertainty goes to zero as the average number of photons goes to infinity. 

From (\ref{lb}), it follows that, for Gaussian states, one may obtain a bound that depends explicitly on the average photon number $N$:
\begin{equation}
 C_Q^{\rm opt}\le C^{\rm max}_Q\equiv \left[2\beta^2+{1\over8N(N+1)}\right]^{-1}\,,
\end{equation}
since, for these states, $\Delta n^{2}\le 2N(N+1).$ 

We compare in Fig.~\ref{fig1} $C^{\rm max}_Q$ with the maximum quantum Fisher information ${\cal F}_Q^{\rm max}$ obtained numerically in Ref.~\cite{Parisprl11} for the best Gaussian probe states with given average photon number $N$. Good qualitative and quantitative agreements between them is observed for $N$ up to 30, which is the range considered in Ref.~\cite{Parisprl11}. On the other hand, our bound, being analytical, allows one to obtain a better insight, for any value of $N$, on the ultimate limit for phase estimation in the presence of phase diffusion. In particular, the saturation of our bound when $16(N\beta)^2\gg1$ is clearly displayed. 

\textit{Summary.}---We have presented in this Letter a variational method to determine the quantum Fisher information by minimizing upper bounds to this quantity, and have given a general prescription to perform this minimization. It bears all the advantages of variational methods, which lead to useful analytical bounds in situations where an exact solution cannot be found analytically. We have applied this method to phase-shift estimation in the presence of phase diffusion and have obtained a nontrivial lower bound to its statistical uncertainty. This bound, which agrees with and goes beyond published numerical results, shows that there exists a constant limit to this uncertainty, which depends only on the strength of the phase diffusion. We believe that the method proposed here might be very useful in determining the fundamental precision limits in quantum metrology in the presence of noise.
 
The authors acknowledge financial support from the Brazilian funding agencies CNPq, CAPES and FAPERJ.  This work was performed as part of the Brazilian National Institute for Science and Technology on Quantum Information.

\end{document}